\documentclass[aps,prl,twocolumn,showpacs,superscriptaddress,groupedaddress]{revtex4-1}
\usepackage{graphicx}

\usepackage{amssymb}
\usepackage{amsmath}
\usepackage{latexsym}
\usepackage{ulem}
\usepackage{color}
\usepackage{dcolumn}
\usepackage{subfigure}
\usepackage[T1]{fontenc}
\usepackage[utf8]{inputenc}
\usepackage[english,french]{babel}
\usepackage[colorlinks=true,linkcolor=blue,citecolor=blue]{hyperref} %
\usepackage{bm}        
\usepackage{amssymb}   

\usepackage[capitalize]{cleveref}

\begin{document}

\title{Giant boost of the quantum metric in disordered one dimensional flat band systems}

\author{G. Bouzerar}
\email[E-mail:]{georges.bouzerar@neel.cnrs.fr}
\affiliation{Université Grenoble Alpes, CNRS, Institut NEEL, F-38042 Grenoble, France}                    
\date{\today}
\selectlanguage{english}
\begin{abstract}
It is a well known fact, that the disorder has its most dramatic effects on the conventional quantum transport in one dimensional systems. In flat band (FB) systems, it is revealed that the conductivity at the FB energy is robust against the disorder and can even be tremendously boosted. Here, the disorder is due to randomly distributed vacancies.
Furthermore, challenging our understanding of the physical phenomena, the giant increase occurs in the limit of low FB states density. The singular behaviour of the quantum metric of the FB eigenstates is found to be at the heart of these unexpected and puzzling features. Additionally, it is shown that the compact localized eigenstates should extend over at least two unit cells to allow a boost.
Our findings should have interesting fallout for other physical systems, and may as well open up engineering strategies to boost the critical temperature in two dimensional superconducting FB materials.
\end{abstract}

\maketitle

Recently, a new class of materials has emerged in the spotlight, the flat band (FB) systems. Their dispersionless bands are at the origin of a plethora of unexpected phenomena \cite{review1,review2}. The quenched kinetic energy promotes the electron-electron interaction and favours the emergence of strongly correlated phases and exotic phenomena, such as fractional quantum Hall states \cite{tang,sun,neupert}, unconventional superconductivity \cite{miyahara,cao,yankowitz}, Wigner crystallization \cite{li,wu1,wu2} and magnetic phases \cite{lin,yin,tasaki}. FBs are as well responsible for an unusual form of quantum electronic transport (QET) as revealed in several studies \cite{gb-fb-paper1,mucciolo,vigh}. It is well established that an infinitesimal amount of disorder destroys the metallic phase (conventional) in one dimensional systems, and leads to the Anderson localization of all the eigenstates. In this work, we address numerically and analytically the QET in disordered one dimensional FB systems with a focus on the conductivity at the FB energy ($\sigma_{fb}$).The singular form of the quantum metric of the disordered FB eigenstates is found to be at the heart of several unforeseen and puzzling features revealed in this work. 

\begin{figure}[t]\centerline
{\includegraphics[width=1\columnwidth,angle=0]{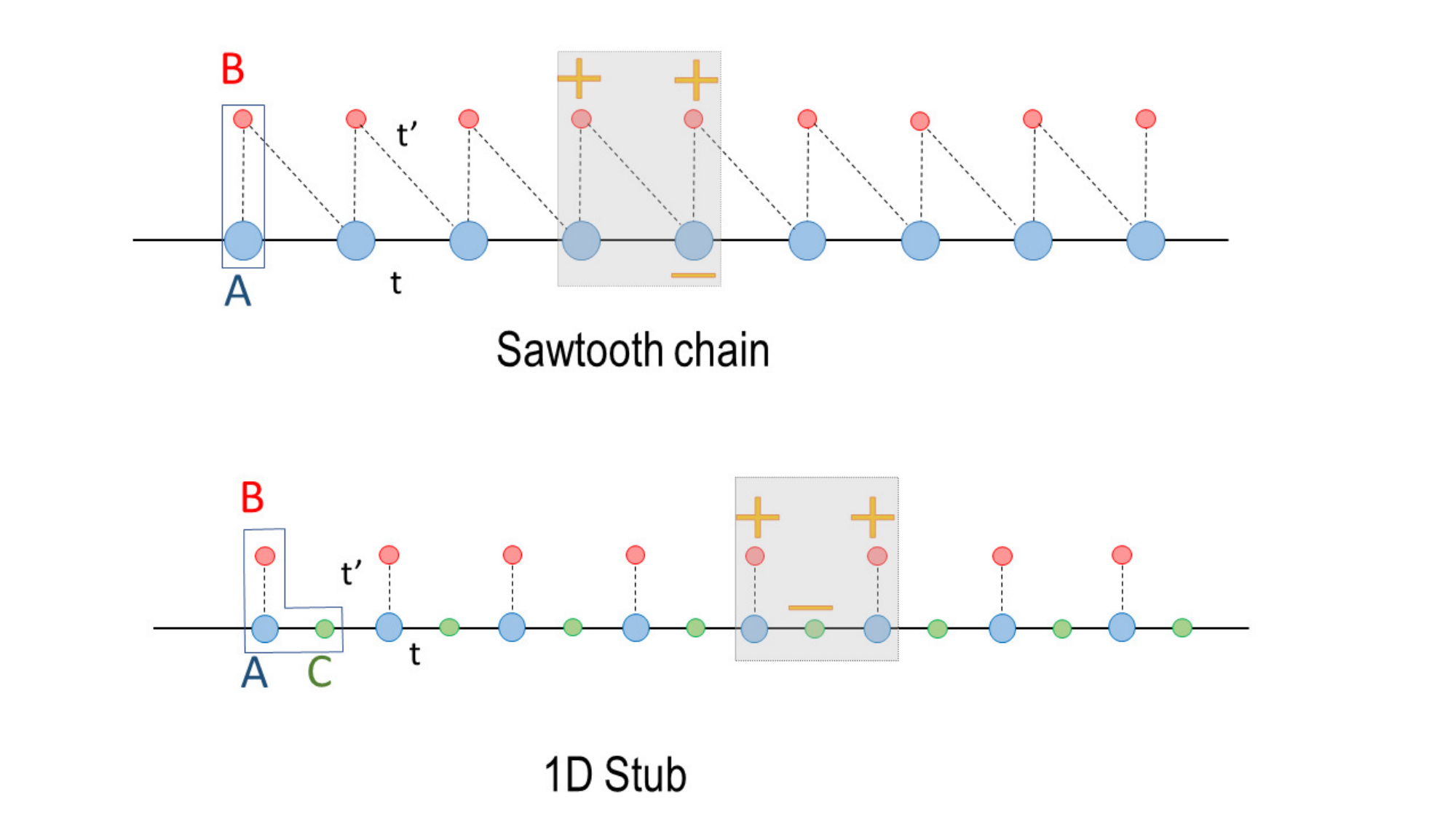}}
\vspace{-0.2cm} 
\caption{(Color online) The sawtooth chain and the stub lattice illustrated.
$t$ is the hopping between nearest neighbour sites along the chain and
$t'$ (dashed line) is the hopping between nearest neighbour pair (B$_i$,A$_i$) in the stub lattice and 
between (B$_i$, A$_{i}$) and (B$_{i}$, A$_{i+1}$) in the sawtooth chain ($i$ is the unit cell index). The grey shaded area represents a typical compact localized flat band eigenstate.
}
\label{fig1}
\end{figure}

We consider two different FB systems, the sawtooth chain (SC) and the stub lattice (SL) as they are illustrated in Fig.~\ref{fig1}. Electrons in these disordered systems are modelled by a tight-binding Hamiltonian,
\begin{eqnarray}
\widehat{H}=-\sum_{\left\langle ij \right\rangle,s} t_{ij} c_{is}^{\dagger}c^{}_{js} + h.c.,
\end{eqnarray}
c$_{is}^{\dagger}$ creates an electron with spin $s$ at site \textbf{R}$_{i}$. The sum runs over the lattice sites, $\left\langle ij \right\rangle$ are pairs for which the hopping $t_{ij}$ is non zero. The disorder is introduced by removing randomly B atoms, since the removal of A (or C) atoms will cut the system into disconnected pieces. The vacancy density is noted $x=\frac{N_{vac}}{N_c}$, $N_{vac}$ is the number of removed atoms and $N_c$ the number of unit cells in the pristine system. In the SL, the FB energy is independent of $t'$, $E_{fb}=0$. In the SC the FB exists only when $t'=\sqrt{2}t$, and $E_{fb}=2t$. From now on, we use this value of $t'$ in the SC and introduce the parameter $\alpha=t'/t$ for the SL. A typical compact localized FB eigenstate (CLS) can be straightforwardly constructed for both lattices. In the SC (resp. SL), a typical CLS is $\vert \psi^{FB}_{i} \rangle = \frac{1}{2}(-\sqrt{2} \vert A_{i} \rangle + \vert B_{i} \rangle + \vert B_{i-1} \rangle)$ (resp. $\vert \psi^{FB}_{i} \rangle= \frac{1}{\sqrt{\alpha^2 +2}}(\vert B_{i} \rangle + \vert B_{i+1} \rangle -\alpha \vert C_{i} \rangle)$), $i$ is the unit cell index. In the SC, the FB is separated from the dispersive band by a large gap of amplitude $\Delta=2t$ whilst in the SL it is tunable $\Delta= \vert \alpha \vert t$. 

\begin{figure}
\includegraphics[width=8cm, height=6.cm]{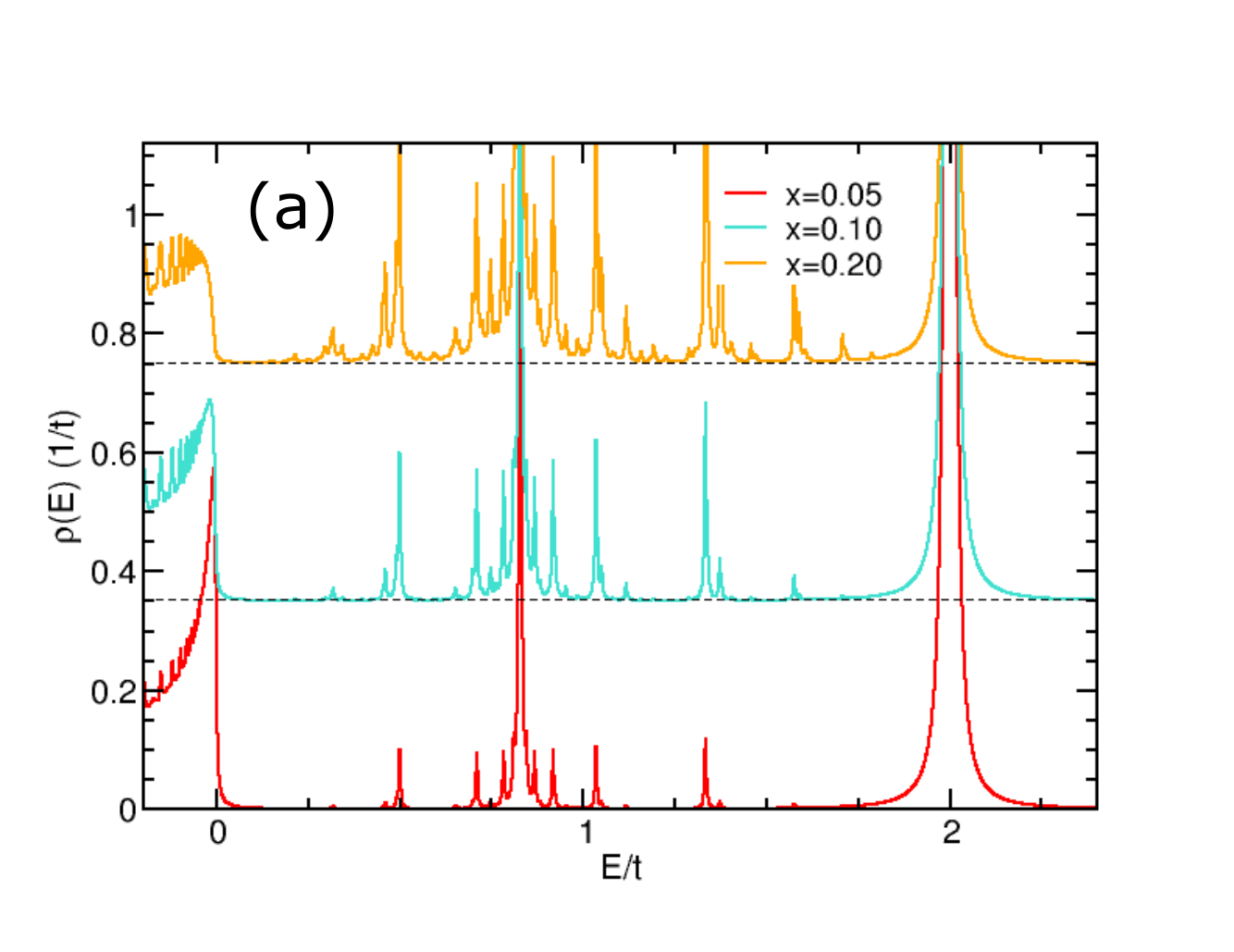}\\
\vspace{-0.4cm} 
\includegraphics[width=8cm, height=6.cm]{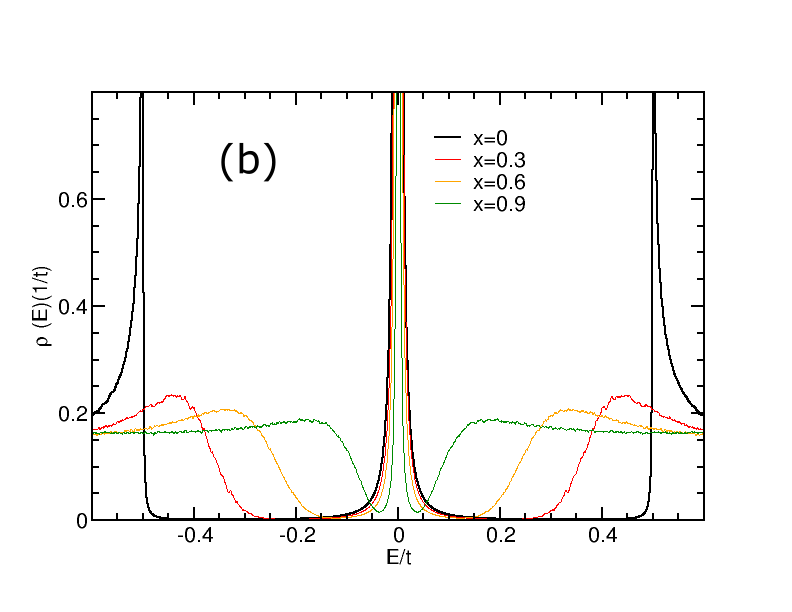}
\vspace{-0.4cm} 
\caption{(Color online) DOS as a function of the energy in the disordered (a) sawtooth chain and (b) stub lattice ($\alpha=0.5$). The vacancy densities ($x$) are indicated in the figures. For more visibility, in (a) two data sets have been shifted upwards.
}
\label{fig2}
\end{figure}
The conductivity along the chain direction is given by the Kubo-Greenwood formula,
\begin{eqnarray}
\sigma(E)=\frac{e^{2}\hbar}{\pi\Omega} \mathrm{Tr} \left[\operatorname{\Im} \widehat{G}(E) 
 \widehat{v}_{x} \operatorname{\Im} \widehat{G}(E) \widehat{v}_{x} \right]. 
\label{eqcond}
\end{eqnarray}
The current operator is $\widehat{v}_{x}=-\frac{i}{\hbar}\left[ \widehat{x} ,\widehat{H} \right]$, where 
$\widehat{x}$ is the position operator ($\widehat{x}=\sum_{is} x_{i} c_{is}^{\dagger}c^{}_{is}$). The Green's function $\widehat{G}(E)=(E+i\eta-\widehat{H})^{-1}$, where $\eta$ mimics a small inelastic scattering rate. $\Omega=N_{c}a$ is the system length (a is the nearest neighbour distance between A sites). To deal with the disorder, the numerical calculations are done using the efficient Chebyshev polynomial Green's function method (CPGF) \cite{mucciolo,weisse,richard,lee} that allows large scale calculations as it requires a modest amount of memory. The calculations $\sigma(E)$ are realized on chains of about $3\,10^{5}$ sites, and an average over at least 100 configurations of disorder is systematically realized. 

Fig.~\ref{fig2} depicts the density of states (DOS), $\rho(E)=-\frac{1}{\pi N_c} \mathrm{Tr} \left[ \operatorname{\Im} \widehat{G}(E)\right]$ in both lattices. In the SC (Fig.~\ref{fig2}\,a), as $x$ increases the weight of the FB states ($E_{fb}=2t$) decreases linearly and localized impurity states start to fill the gapped region. We notice multiple peaks emerging in the dispersive part of the DOS, their density increases significantly as $x$ increases. $\rho(E)$ in the disordered SL is shown in Fig.~\ref{fig2}\,b for $\alpha=0.5$. A first glance reveals a rather different picture from what is observed in the SC. As $x$ increases, the gap reduces, the divergence at the upper (resp. lower) edge of the valence (resp. conduction) band is removed, $\rho(E)$ becomes smoother and flatter in this region. As it is expected, the number of FB states is $(1-x)N_{c}$, since each vacancy destroys a CLS. Remark that in the disordered system, we still use "flat band" to characterize the degenerate $E=E_{fb}$ eigenstates. Indeed, if we calculate the spectral function $A(\textbf{k},E)=-\frac{1}{\pi} \langle \Im G(\textbf{k},\omega) \rangle$, $\langle ..\rangle$ is the average over the disorder and $G(\textbf{k},E)$ is the Fourier transform of $G_{ij}(E)$, we would find a flat band at $E=E_{fb}$.

\begin{figure}[t]\centerline
{\includegraphics[width=1\columnwidth,angle=0]{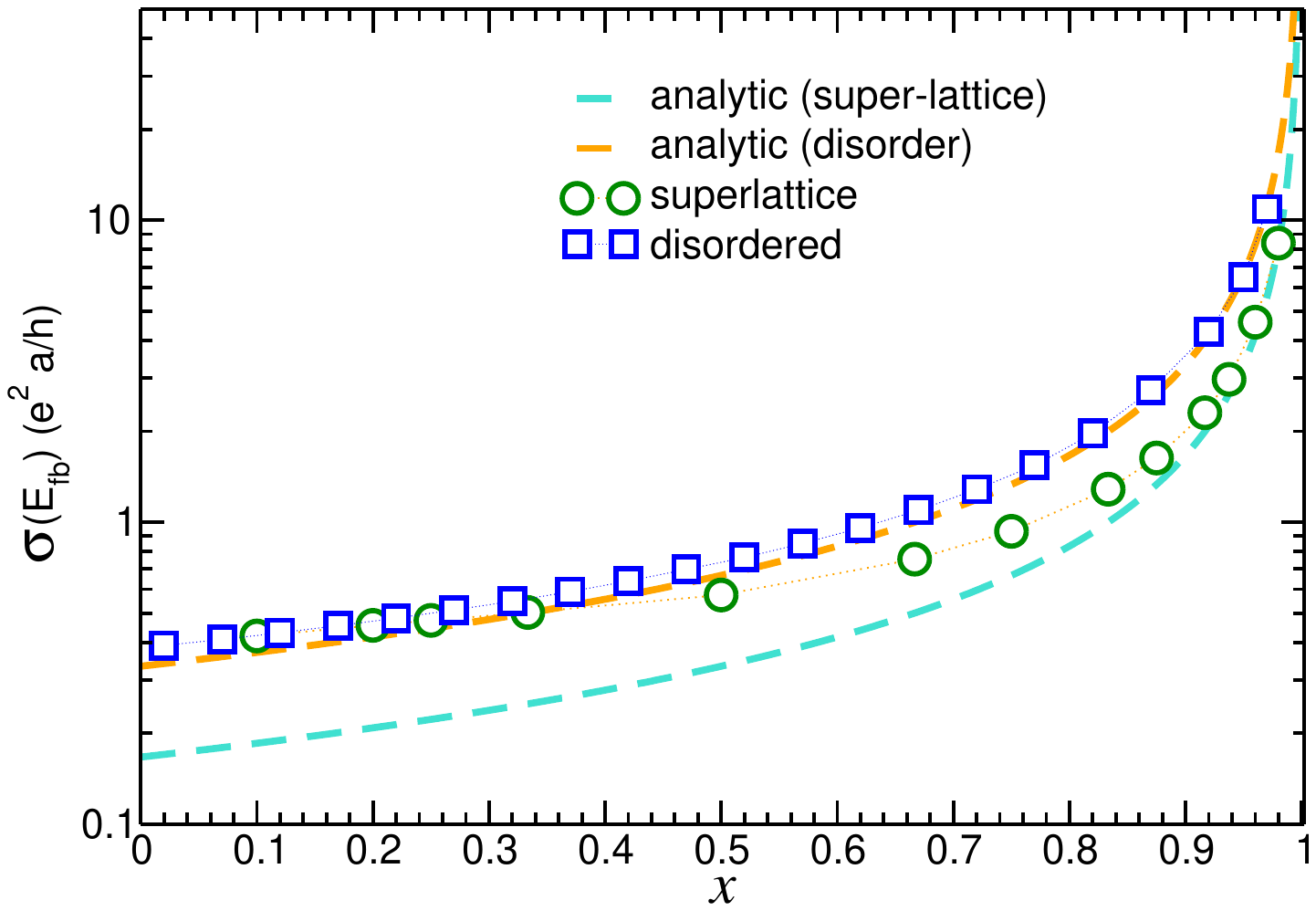}}
\vspace{-0.6cm}
\caption{(Color online) 
FB conductivity in the sawtooth chain as a function of $x$.
The numerical calculations are performed for vacancies (i) randomly distributed ('disorder') and (ii) organized ('superlattice'). The dashed lines correspond to the analytical approach described in the main text.} 
\label{fig3}
\end{figure} 
We now discuss the impact of the disorder on the conductivity at the FB energy. We recall that the intraband (Drude) term does not contribute because of the vanishing group velocity, $\sigma_{fb}$ reduces to the interband contribution \cite{gb-fb-paper1}. Fig.~\ref{fig3} depicts $\sigma_{fb}$ in the SC, as a function of $x$. Here, the data correspond to the limit of vanishing $\eta$. For $x < 0.80$, $\eta$ has a negligible effect, because of the large interband gap. As $x\rightarrow 1$, the gap region gets progressively filled, the $\eta$-dependence becomes stronger requiring a careful analysis. We first focus on the clean SC ($x=0$), for which the analytical calculations leads to $\sigma_{fb}=\frac{2}{3\sqrt{3}}\,\sigma_{0}$ where $\sigma_{0}=\frac{e^{2}}{h}a$. As $x$ increases, $\sigma_{fb}$ increases. One would have expected a reduction instead
since the FB states density reduces. As $x$ increases further, $\sigma_{fb}$ increases much more rapidly and eventually diverges as $x \rightarrow 1$, which is even more intriguing. Indeed, for $x=1$, the system reduces to a trivial 1D chain, the conductivity is purely of Drude type (intraband) and it is easy to show that $\sigma(E)= \sqrt{4t^{2}-E^{2}}\, \frac{\sigma_{0}}{2\eta}$. Thus, $\sigma(E=2t)=0$, since no holes are available (or electrons for $E=-2t$) in the chain. Consequently, it would have been reasonable to expect a decay of the conductivity as $x \rightarrow 1$. The origin of these unexpected features is clarified below. To evaluate the impact of the disorder, we consider a SC where the remaining B atoms are ordered (superlattice). In other words B atoms are located at $x_n=\frac{n}{1-x} a$, where $n=0,1,...$, and $x$ is chosen appropriately ($\frac{1}{1-x}$ should be integer).

The data are plotted in Fig.~\ref{fig3} as well. For $x \ll 1$, the conductivity in the disordered and in the ordered SC coincide almost perfectly with each other, as it could have been anticipated. Astonishingly, as $x$ increases further ($x > 0.3$), $\sigma_{fb}$ becomes smaller in the ordered than in the disordered system. Finally, as $x \rightarrow 1$, $\sigma_{fb}$ diverges as well in the superlattice but it is approximately half of that of the disordered SC. Thus, as unusual as it may be, the disorder enhances the conductivity.

Within an analytical approach, our goal is to clarify the origin of the unusual features brought to light in our numerical calculations. Here, we only summarize the main steps of the procedure, the full details are available in the Appendices (A,B and C). It is more convenient to start from the limit of high vacancy concentration, and introduce $y=1-x$ the density of B atoms distributed randomly in the chain, we assume $y \ll 1$. We define a configuration of disorder by the position in the SC of the $N_B$ B-atoms ($N_B=y\times N_{c}$): $(B_{p_{0}},B_{p_{1}},B_{p_{2}},...,B_{p_{N_{B}-1}})$ where $p_0 < p_1 < ....< p_{N_{B}-1}$. For each pair ($B_{p_{m-1}}, B_{p_{m}}$), using a linear combination of the CLS states of the clean chain, one can construct $\vert \psi^{FB}_m \rangle$ a FB eigenstate of the disordered SC, where $\langle A_i \vert \psi^{FB}_m \rangle$ is non zero only for the sites located between $B_{p_{m-1}}$ and $B_{p_{m}}$ (see Appendix A). We find,
$\vert \psi^{FB}_m \rangle = \vert \bar{\psi}_{p_{m-1}p_{m}} \rangle / \langle \bar{\psi}_{p_{m-1}p_{m}} \vert \bar{\psi}_{p_{m-1}p_{m}} \rangle$ where,
\begin{eqnarray}
\vert \bar{\psi}_{p_{m-1}p_{m}} \rangle &=&-e^{-i\varphi} (\vert B_{p_{m-1}} \rangle + (-1)^{p_{m}}e^{i(p_{m})\varphi} \vert B_{p_{m}} \rangle)  \nonumber \\
&+& \sqrt{2}\sum^{p_{m}}_{l=p_{m-1}+1} e^{i(l-1)\varphi} (-1)^{l-1} \vert A_l \rangle,
\label{eqfbb}
\end{eqnarray}
where the phase $\varphi= \frac{2\pi}{N_c} \frac{\Phi}{\phi_0}$, $\Phi$ being a magnetic flux threading the sawtooth ring ($\phi_0=\frac{h}{e}$) introduced to allow the calculation of the conductivity. These disordered CLS are linearly independent, but not orthogonal to each other. Since our main focus is $y\ll 1$, we can neglect the overlap between them. Starting from the Kubo-Greenwood formula and using the fact that $\hat{v}_{x}=-\frac{a}{\hbar}\dfrac{\partial \hat{H}}{\partial \varphi}$, in the disordered SC, we find, 
\begin{eqnarray}
\sigma_{fb}=2 y \, \langle g^{m}_{\varphi \varphi}\rangle \,\sigma_0,
\label{eqsigcb}
\end{eqnarray}
where $\langle .. \rangle$ means average over the FB eigenstates and $g^{m}_{\varphi \varphi}$ is given by,
\begin{eqnarray}
g^{m}_{\varphi \varphi} = \langle \partial_{\varphi} \Psi_{m}^{FB} \vert \partial_{\varphi} \Psi_{m}^{FB} \rangle - \vert \langle \partial_{\varphi} \Psi_{m}^{FB} \vert \Psi_{m}^{FB} \rangle \vert^{2}.
\label{qm}
\end{eqnarray}
This is the quantum metric (QM) associated to the FB eigenstate $\vert \Psi_{m}^{FB} \rangle$. The concept of QM has been originally introduced in Ref.~\cite{qm1} and discussed in various context in Ref.~\cite{qm2,qm3,qm4,qm5}. In our one dimensional space spanned by $\varphi$, the QM defines a gauge invariant distance between nearby states $\vert \Psi_{m}^{FB} (\varphi) \rangle$ and $\vert \Psi_{m}^{FB} (\varphi+d\varphi) \rangle$, that reads $ds^{2}=1-\vert \langle \Psi_{m}^{FB} (\varphi) \vert \Psi_{m}^{FB} (\varphi+d\varphi) \rangle \vert^{2} = g^{m}_{\varphi \varphi} d\varphi^{2}$. Thus, Eq.(\ref{eqsigcb}) gives a geometric interpretation of the FB conductivity in the disordered system. This is similar to the geometric contribution to the superfluid weight found in superconducting non disordered FB systems \cite{peotta,julku}. 
Eq.(\ref{eqsigcb}) indicates that $\sigma_{fb}$ is proportional to the FB states density, which is conceivable and expected, but as it will be seen the dependence of the QM on $y$ is critical. Notice that Eq.(\ref{eqsigcb}) has been also recently derived independly in Ref.~\cite{Mitscherling} for non disordered systems.

Using the CLS expression given in Eq.(\ref{eqfbb}), we obtain $g^{m}_{\varphi \varphi} \approx \frac{1}{12} (p_{m}-p_{m-1})^{2}$. As it shown in the Appendix B, the average of QM over the FB eigenstates is,
\begin{eqnarray}
\left\langle g^{m}_{\varphi \varphi}\right\rangle = \frac{1}{6y^{2}}.
\label{qmdis}
\end{eqnarray}
Thus, within our analytical approach, the FB conductivity has a simple form, $\sigma_{fb}=\frac{1}{3y} \sigma_{0}$. This explains the origin of the divergence found in the dilute regime. The associated data correspond to the orange dashed line in Fig.~\ref{fig3}. As it can be observed, the agreement with the numerical calculation is excellent for $x > 0.8$ and even surprisingly very good down to $x=0$. Indeed, by construction, the analytical expression is valid for $y \ll 1$, the overlap between the constructed CLS eigenstates is negligible is regime only. For the clean SC ($y=1$), the analytical result corresponds to approximately 87\% of the exact value. We recall as well that our numerical calculations have revealed that when B atoms are distributed regularly, the conductivity is approximately half of that of the disordered system. In the ordered system, we find $\langle g^{m}_{\varphi \varphi}\rangle = \frac{1}{12y^{2}}$, which is exactly half of that of the disordered system as it is given in Eq.(\ref{qmdis}). This definitely clarifies why the disorder enhances the FB conductivity.
\begin{figure}[t]\centerline
{\includegraphics[width=1\columnwidth,angle=0]{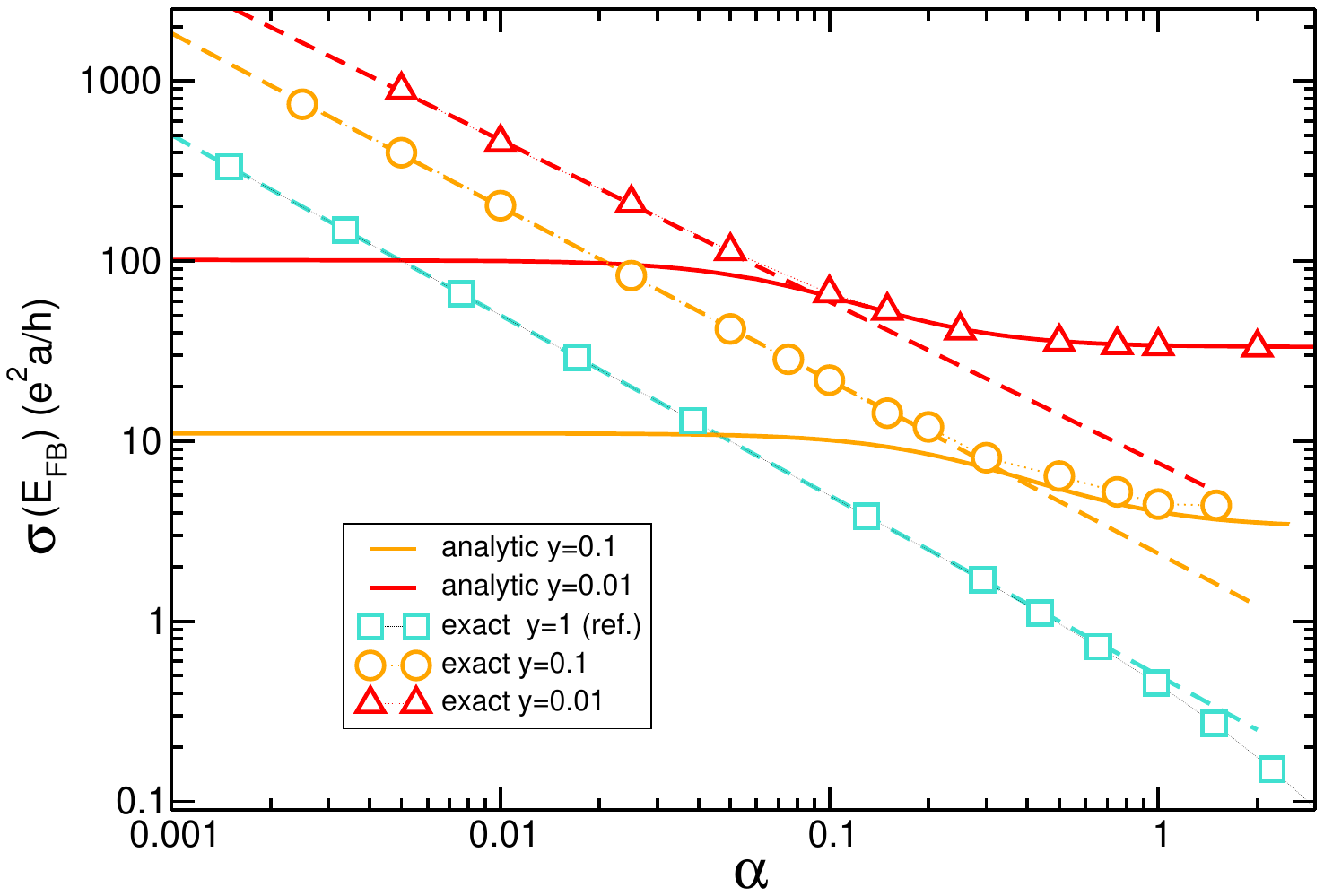}}
\vspace{-0.5cm}
\caption{(Color online) 
FB conductivity as a function of $\alpha$ in the disordered stub lattice, for different values of $y$ ($y=1-x$). 'exact' corresponds to the numerical calculations. The continuous lines are the analytical calculations (see text) and the dashed lines simply a guide to the eye.
} 
\label{fig4}
\end{figure} 
Notice that there is an alternative way to express the FB conductivity in terms of the mean spread of FB eigenstates. As it is shown in the Appendix B, Eq.(\ref{eqsigcb}) can be written,
\begin{eqnarray}
\sigma_{fb}=2 y \, \langle L_{m}^{2} \rangle \,  \sigma_{0},
\label{eqsigcd}
\end{eqnarray}
where $L_{m}= \left[ \langle \Psi_{m}^{FB} \vert (\hat{x} - \langle \hat{x} \rangle)^{2} \vert \Psi_{m}^{FB} \rangle \right]^{1/2}$, is the mean spread of $\vert \Psi_{m}^{FB} \rangle$. The relation between the QM and the spread of the eigenstates has been introduced in the context of maximally localized Wannier \cite{marzari1,marzari2}. The diverging conductivity in the dilute regime results from the $1/y^{2}$ dependence of the spread of the FB eigenstates that overcompensates the low FB density.

To show the universal character of findings, we address the disorder effects in the SL. Since the SL offers the degree of freedom to tune the gap (or $\alpha$) without destroying the FB, we consider the combined effects of $y$ and $\alpha$ on $\sigma_{fb}$. In Fig.~\ref{fig4}, $\sigma_{fb}$ is plotted as a function of $\alpha$ for 3 different values of $y$. To start, we discuss the case of the clean system for which the exact analytical calculations can be realized. As it is shown in the Appendix C, for $y=1$, $\sigma_{fb}=\dfrac{\sigma_{0}}{\vert \alpha \vert (4+\alpha^{2})^{1/2}}$, it is depicted in Fig.~\ref{fig4} by the blue symbols. $\sigma_{fb}$ scales as $1/\alpha$ as $\alpha \rightarrow 0$. But, as in the SC, $\alpha=0$ corresponds the trivial 1D chain, where the B atoms are disconnected. The transport is of intraband nature in this limiting case, and $\sigma_{E=0}=\frac{t}{\eta}\sigma_{0}$. This implies a transition when $\alpha$ is switched on, the QET changes from intraband to interband type. We now switch to the effects of removing B atoms randomly. For a given $\alpha$, as $y$ decreases, one observes a strong increase of $\sigma_{fb}$ with respect of that of $y=1$.
For example, for $\alpha=1$, the conductivity jumps from $0.4 \, \sigma_{0}$ for $y=1$ to $5 \, \sigma_{0}$ when 90\% of B atoms have been removed, or equivalently when the FB states density has been divided by 10. If $y$ reduces further, $\sigma_{fb}$ increases even more and reaches $\sigma_{fb} \approx 36 \,\sigma_{0}$ for $y=0.01$. Thus, one finds a spectacular boost of $\sigma_{fb}$ on two orders of magnitude when 99\% of the B atoms have been removed, or when the density of FB states represents 1\% only of that of the pristine SL. We discuss now the effect of reducing $\alpha$ for $y=0.1$ and $y=0.01$.
In both cases, a crossover in the vicinity of $\alpha \approx \sqrt{y}$ is visible. Our numerical data show that $\sigma_{fb}$ weakly depends on $\alpha$ when $\alpha > \sqrt{y}$ (far from the crossover). In contrast, when $\alpha < \sqrt{y}$ it strongly increases as $\alpha$ reduces. Using a fit of the form $1/y^{\beta}$, we find $\beta \approx 0.96$ for $y=0.1$ and $\beta \approx 0.88$ for $y=0.01$. As it has been done for the disordered SC, we now analyse the vacancy effects analytically. For a fixed configuration of disorder, we first construct the FB eigenstates basis and then we calculate QM associated of these states. The details are available in the Appendix C. We will not give the general expression of $\sigma_{fb}$ in terms of $\alpha$ and $y$, since it is rather complicated but it is still analytical.
The analytically calculated $\sigma_{fb}$ corresponds to the continuous lines in Fig.~\ref{fig4}. As it has been found numerically, we observe as well a clear crossover $\alpha \approx \sqrt{y}$.
For $\alpha > \sqrt{y}$, the agreement between the analytical results and the exact numerical calculations is very good for $y=0.1$ and even excellent for $y=0.01$. On the other hand, when $\alpha < \sqrt{y}$, the analytical calculation reveals a saturation of $\sigma_{fb}$ as $\alpha$ decreases, contradicting the diverging behaviour found in the exact calculations. The question which arises is why do these calculations disagree so drastically when $\alpha < \sqrt{y}$? As explained before, the procedure used for the analytical calculations is valid only when the overlap between the constructed FB eigenstates can be neglected. A FB state of length $m$ is of the form (see Appendix C), $\vert \psi_{1,m+1}\rangle = \frac{1}{\sqrt{m\alpha^{2}+2}} (\vert B_1  \rangle +(-1)^{m-1} \vert B_{m+1} \rangle) \nonumber \\  
+ \frac{\alpha}{\sqrt{m\alpha^{2}+2}} \sum_{k=1}^{m} (-1)^{k}  \vert C_{k} \rangle$. Thus, the ratio of the weight on type B atoms to that on type C atoms is $\frac{2}{m\alpha^2}$. Hence, the overlap between the non orthogonal FB eigenstates is negligible if $m\alpha^2 \gg 1$, or equivalently when $\alpha^2 \gg y$. This clarifies the presence of a crossover and why the agreement between exact and analytical results is found for $\alpha > \sqrt{y}$ only. There is no simple way to derive an analytical expression of the conductivity for $\alpha < \sqrt{y}$. It would require, a systematic orthogonalization of the FB eigenstates using for instance a Gram-Schmidt procedure which would not lead to a simple analytical form for $\sigma_{fb}$.
We believe that our findings, could be addressed experimentally. Recently, using STM to manipulate individual vacancies in a chlorine monolayer on Cu(100) it has been possible to construct various one-dimensional(1D) lattices with engineered flat band \cite{huda}.

\begin{figure}[t]\centerline
{\includegraphics[width=1\columnwidth,angle=0]{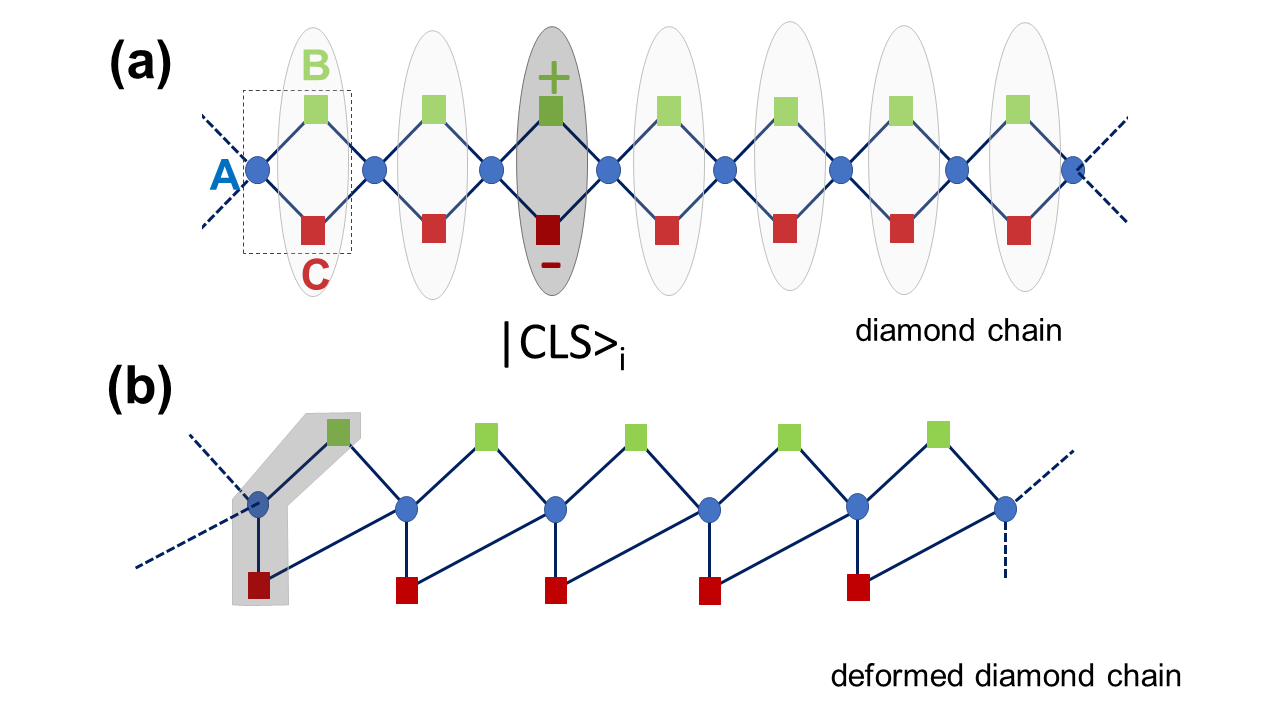}}
\vspace{-0.2cm} 
\caption{(Color online) Illustration of the standard diamond chain and the deformed one which is obtained by shifting the C atoms. The grey shaded area correspond to the CLS eigenstates.}
\label{fig5new}
\end{figure}

It has been shown recently, that the superfluid weight, which has also previously been related to the QM, should be independent of the choice of the orbital positions. Since the QM depends on the position of the orbitals in the unit cell, the relevant quantity for superconductivity in isolated flat bands is the minimal quantum metric \cite{huhtinen,peottabis}.
Here, because Eq.(\ref{eqsigcb}) is valid for any position of the orbitals in the unit cell, it implies that the FB conductivity depends on the orbital positions as well.
For instance, in the clean sawtooth chain case considered in this study and illustrated in Fig.\ref{fig1}, $\sigma_{fb}=\frac{2}{3\sqrt{3}} \,\sigma_{0} \approx 0.385\,\sigma_{0}$. For the symmetric sawtooth chain where the distance between B atoms and their nearest neighbours A is identical, we find a smaller value $\sigma_{fb}=\frac{3+\sqrt{3}}{18}\,\sigma_{0} \approx 0.263\,\sigma_{0}$. This later case corresponds to the minimal QM or equivalently to the minimal conductivity.

We propose now to shed light on the importance that CLS states extend over at least two unit cells to obtain a boost of the FB conductivity. For that purpose, we consider the one dimensional diamond lattice as it is illustrated in Fig.\ref{fig5new} and for which the CLS eigenstates occupy a single unit cell (shaded grey area in the illustration).
As a consequence, the overlap between CLS states is zero. The set of CLS eigenstates 
$\vert CLS \rangle_i=\frac{1}{\sqrt{2}}(\vert i,B \rangle -\vert i,C \rangle )$ where $i$ labels the unit cell and $i = 1,...,N$ is a FB basis. Thus, for the disordered chain the FB basis is obtained by simply removing the CLS states associated to the vacancy sites. Hence, the average of the QM does not depend on $y$, thus the FB conductivity decays linearly as $y$ is reduced. Consider the case of the diamond chain as it is illustrated in \ref{fig5new})(a), for which we find $\sigma_{fb}=0$ for any value of $y$. In contrast, for the disordered deformed diamond chain (Fig.\ref{fig5new}(b)), the calculation of the QM leads to $\sigma_{fb}=0.125 y \,\sigma_{0}$.
This example nicely illustrates the fact that the condition that CLS states extend over at least two unit cells is crucial to boost the FB conductivity.

To conclude, it is revealed that the dilution of FB eigenstates can lead to a giant boost of the flat band conductivity. 
At the origin of this unexpected and counter-intuitive physical phenomenon is the diverging behaviour of the quantum metric of the FB states. It is also shown that the condition that CLS eigenstates extend over at least two unit cells is crucial. The physics highlighted in this work is general and not restricted to one dimensional systems. 
As it has been shown recently, the quantum metric plays a key role on the amplitude of the critical superconducting temperature in FB systems, we argue that our findings may as well open up strategies to engineer high-Tc materials. The STM manipulation of adatoms on the surface of two dimensional materials, or the intercalation of atoms between multilayer compounds could be interesting pathways. The promising candidates could be identified with the efficient support of first principles studies.

\begin{acknowledgments}
We thank M. Nunez-Regueiro, P. Rodière for interesting discussions. We thank as well one of the referees to bring to our attention three recent and relevant publications \cite{huhtinen,peottabis,Mitscherling}.
\end{acknowledgments}

\subsection{Appendix A : Construction of the FB eigenstates for the disordered sawtooth chain}

\renewcommand{\theequation}{A.\arabic{equation}}

\setcounter{equation}{0}


The aim of this paragraph is to show the procedure used to construct the FB eigenstates basis in the disordered sawtooth chain (SC).

To allow the analytical calculation of the conductivity, we consider a SC ring threaded by magnetic flux $\Phi$.
This results in the well known Peierls substitution in the tight binding Hamiltonian: $t_{ij} \rightarrow t_{ij}exp(-i\frac{e}{\hbar}\int_{i}^{j}\textbf{A}.\textbf{dl})$ where $\oint\textbf{A}.\textbf{dl}=\Phi$, and because we choose $\textbf{A}=A_{x}\textbf{e}_{x}$ uniform $A_{x}N_ca=\Phi$, $N_c$ is the number of unit cell (system size). We recall that the current operator is then given by, $\hat{j}_{x}=-\dfrac{\partial \hat{H}}{\partial A_{x}}$.
In the clean case, a typical flux dependent FB-eigenstate in the sawtooth chain is given by,
\begin{eqnarray}
\vert FB^{0} \rangle_{i}=\frac{1}{2}(\sqrt{2}\vert A_i \rangle -e^{-i\varphi} \vert B_{i-1} \rangle -\vert B_{i} \rangle),
\end{eqnarray}
where $\varphi=\frac{2\pi}{N_c} \frac{\Phi}{\phi_{0}}$, and $\phi_{0}=h/e$ is the quantum flux unit.

In this section we focus in the limit of large concentration of vacancies. The remaining B atoms are very dilute, thus far from each other, we denote $y=1-x$ ($y \ll 1$) their concentration. We define a configuration of disorder by the position of the $N_{B}$ B-sites (where $N_{B}=y\times N_{c}$): $(B_{m_{0}},B_{m_{1}},B_{m_{2}},...,B_{m_{N_{B}-1}})$, where the position of the B atoms are organized in increasing order $m_{0} < m_{1} < ....< m_{N_{B}-1}$. For such a configuration one can build the set of $N_{B}$ FB eigenstates of the disordered Hamiltonian.
We denote $\vert \psi^{FB}_{m_{k-1} m_{k}} \rangle$ the $E=E_{fb}$ eigenstate that has non zero components between $B_{m_{k-1}}$ and $B_{m_{k}}$ where $k=1,..,N_{B}-1$. Because of the periodic boundary conditions, the missing $N_{B}$-th FB state is obtained for the pair $(B_{m_{N_{B}-1}},B_{m_{0}})$.

\begin{figure}[t]\centerline
{\includegraphics[width=1\columnwidth,angle=0]{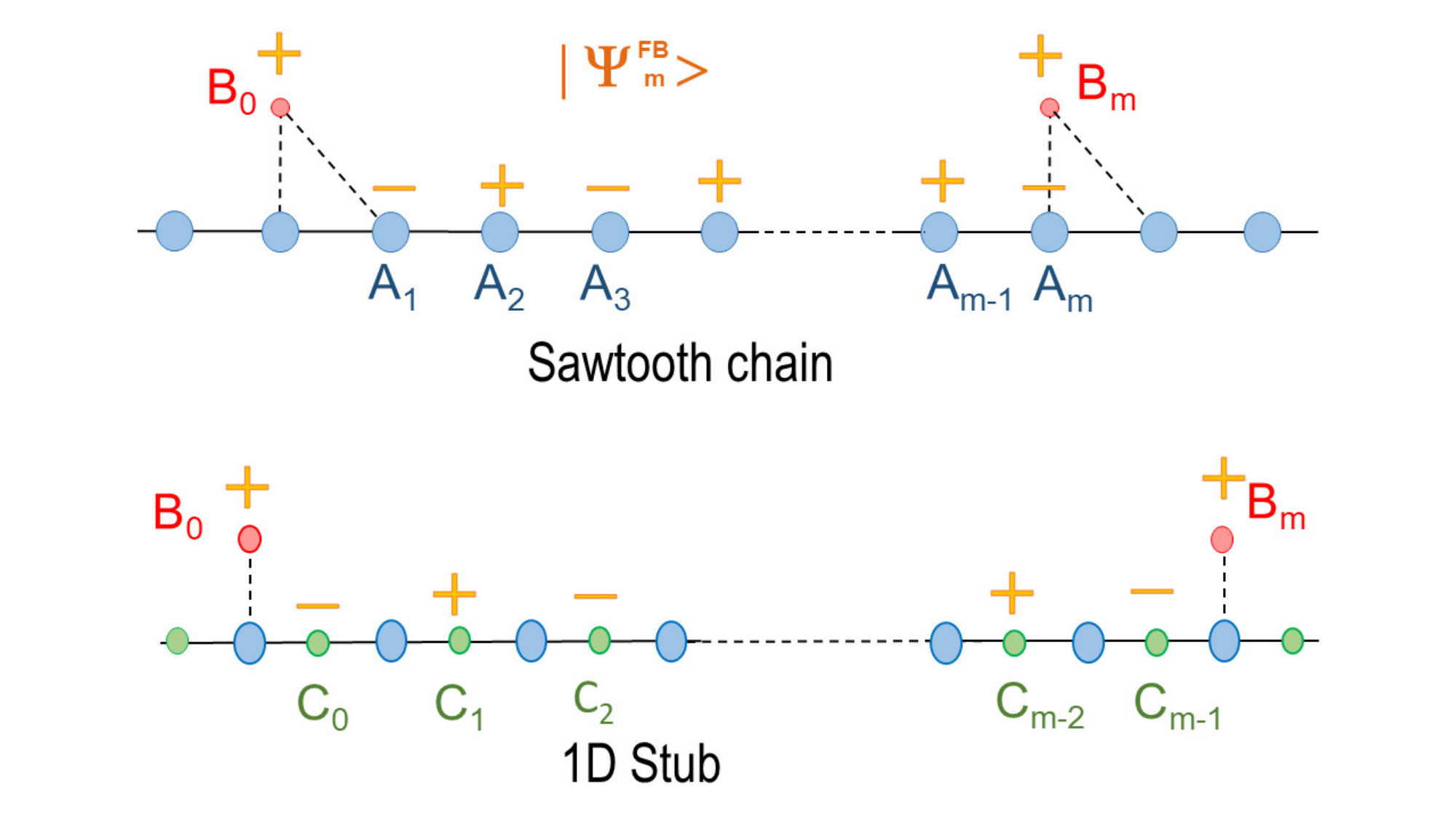}}
\caption{(Color online) 
Structure of $\vert \Psi_{m}^{FB} \rangle$, a typical FB eigenstates used for the analytical calculation of the FB conductivity in the disordered sawtooth chain and in the disordered stub lattice.
} 
\label{fig5}
\end{figure}

Let us first start with the case of a single pair of B sites located respectively at $B_{m_{0}}$ and $B_{m_{1}}$, the other B atoms have been removed and we assume that this pair of B atoms are far away from each other as it is illustrated in Fig.~ \ref{fig5}. First, from the CLS eigenstates of the pristine Hamiltonian ($\hat{H_0}$), we construct an eigenstate which has vanishing components on the B atoms located between $B_{m_{0}}$ and $B_{m_{1}}$, 
\begin{eqnarray}
\vert \bar{\psi}_{m_{0}m_{1}} \rangle = \sum^{m_{1}}_{p=m_{0}+1} e^{i(p-1)\varphi} (-1)^{p-1} \vert FB^{0} \rangle_{p}.
\end{eqnarray}
This state can be re-written,
\begin{eqnarray}
\vert \bar{\psi}_{m_{0}m_{1}} \rangle &=&-e^{-i\varphi}\vert B_{m_{0}} \rangle + (-1)^{m_{1}}e^{i(m_{1}-1)\varphi} \vert B_{m_{1}} \rangle  \nonumber \\
&+& \sqrt{2}\sum^{m_{1}}_{p=m_{0}+1} e^{i(p-1)\varphi} (-1)^{p-1} \vert A_p \rangle.
\label{eqfb1}
\end{eqnarray}
It is an eigenstate of $\hat{H}=\hat{H_0}-\delta \hat{H}$, where $\delta \hat{H}$ corresponds to all removed (A,B) hoppings located between the pair ($B_{m_{0}}$,$B_{m_{1}}$). One can easily check that $\delta \hat{H} \vert \bar{\psi}_{m_{0}m_{1}} \rangle = 0$. Since this state is a linear combination of the CLS eigenstates of $\hat{H_0}$, then it is also an eigenstate of the disordered Hamiltonian $\hat{H}$. The corresponding normalized eigenstate is defined by 
\begin{eqnarray}
\vert \psi_{m_{0}m_{1}}\rangle=\frac{\vert \bar{\psi}_{m_{0}m_{1}} \rangle}{\sqrt{2(d_{1,0}+1)}}. 
\label{eqfb2}
\end{eqnarray}
where, $d_{i,i-1}=m_{i}-m_{i-1}$. Hence, for a given configuration of the disorder (position of B sites) given by $(B_{m_{0}},B_{m_{1}},B_{m_{2}},...,B_{m_{N_{B}-1}})$, we can construct the set of $N_B$ ($N_B=y N_c$) eigenstates with energy $E_{fb}$ following this procedure.

These states are linearly independent but not orthogonal to each other. Indeed, two successive eigenstates overlap at one of the B sites. For instance the overlap $\langle \psi_{m_{i}m_{i+1}} \vert \psi_{m_{i+1}m_{i+2}} \rangle$, is given by $\frac{1}{2\sqrt{(d_{i+1,i}+1)(d_{i+2,i+1}+1)}}$ , where we ignore the phase factor.
In the dilute limit $\langle \psi_{m_{i}m_{i+1}} \vert \psi_{m_{i+1}m_{i+2}} \rangle$ is of the order of $y/2$, and thus one can ignore the small overlap between our constructed eigenstates. Within this approximation, for our configuration of disorder, the basis of FB eigenstates is ($\vert \psi_{m_{0}m_{1}} \rangle$,$\vert \psi_{m_{1}m_{2}}\rangle$,.....,$\vert \psi_{m_{N_{B}-1}m_{N_{B}}}\rangle$,$\vert \psi_{m_{N_{B}}m_{0}}\rangle$).

\subsection*{Appendix B : Flat band conductivity and quantum metric in the sawtooth chain}

\renewcommand{\theequation}{B.\arabic{equation}}

\setcounter{equation}{0}

The aim of this section is to show the relation between the conductivity at the flat band energy and the quantum metric in the case of the disordered sawtooth chain. The full basis of the disordered sawtooth eigenstates is defined by,
$(\left\lbrace \vert \Psi_{j}^{FB} \rangle \right\rbrace_{j}, \left\lbrace \vert \Phi_{l} \rangle \right\rbrace_{l})$ where 
$\vert \Psi_{j}^{FB} \rangle = \vert \psi_{m_{j}m_{j+1}}\rangle$ ($j=1,2,...,yN_c$) are the FB eigenstates as defined in the previous section, and $\vert \Phi_l \rangle$ correspond to the dispersive eigenstates with eigenenergy $E_l \ne E_{fb}$ where $l=1,...,N_c$.

Starting from the Kubo-Greenwood formula as it is given in the main text, we can re-express the conductivity at $E=E_{fb}$,
\begin{eqnarray}
\sigma_{fb}=\frac{2e^{2}a}{h N_c}  \sum_{j,l} \vert \langle \partial_{\varphi} \Psi_{j}^{FB} \vert \Phi_l \rangle \vert^{2}\frac{\bar{E_l}^{2}}{\bar{E_l}^{2}+\eta^{2}},
\label{eqsigb}
\end{eqnarray}
where $\bar{E_l}=E_l-E_{fb}$. Notice that we have used the relation $\partial_{\varphi} \hat{H} \vert \Psi_{j}^{FB} \rangle = (E_{fb} - \hat{H}) \vert \partial_{\varphi} \Psi_{j}^{FB} \rangle$. For $\eta$ small enough, in other words, smaller than $\delta E=\min(\bar{\vert E_l \vert})$, we can replace $\frac{\bar{E_l}^{2}}{\bar{E_l}^{2}+\eta^{2}}$ by 1. Notice that the systems that have studied numerically indicate a gap between the FB and the dispersive eigenstates. Hence, by inserting in Eq.(\ref{eqsigb}) the relation $\sum_{l} \vert \Phi_l \rangle \langle \Phi_l \vert = 1-\sum_{j} \vert \Psi_{j}^{FB} \rangle \langle  \Psi_{j}^{FB} \vert$, the FB conductivity becomes,
\begin{eqnarray}
\sigma_{fb}=\frac{2e^{2}a}{h N_c} \sum_{j} \left[ \langle \partial_{\varphi} \Psi_{j}^{FB} \vert \partial_{\varphi} \Psi_{j}^{FB} \rangle - \sum_{k}\vert \langle \partial_{\varphi} \Psi_{j}^{FB} \vert \Psi_{k}^{FB} \rangle \vert^{2} \right]. \nonumber \\
\end{eqnarray}
The overlap $C_{jk}=\langle \partial_{\varphi} \Psi_{j}^{FB} \vert \Psi_{k}^{FB} \rangle$ is non zero only for $j=k$ and $j=k\pm 1$. In the same way that we could neglect the overlaps $ \langle \Psi_{k}^{FB} \vert \Psi_{k\pm 1}^{FB} \rangle$
because they are of the order of $y/2 \ll 1$, we ignore as well $C_{j,j\pm 1}$, and keep only the $C_{jj}$. We obtain,
\begin{eqnarray}
\sigma_{fb}=2 y \, \langle g^{m}_{\varphi \varphi}\rangle \,\sigma_0,
\label{eqsigc}
\end{eqnarray}
where $\langle .. \rangle$ means average over the FB eigenstates, and $\sigma_{0}=\frac{e^{2}}{h} a$.
$g^{m}_{\varphi \varphi}$ is the quantum metric associated to the disordered FB eigenstate $\vert \Psi_{m}^{FB} \rangle$. The concept of quantum metric has been originally introduced in Ref.~\cite{qm1} and discussed in various context in Ref.~\cite{qm2,qm3,qm4,qm5}. 
Here it reads,
\begin{eqnarray}
g^{m}_{\varphi \varphi} = \langle \partial_{\varphi} \Psi_{m}^{FB} \vert \partial_{\varphi} \Psi_{m}^{FB} \rangle - \vert \langle \partial_{\varphi} \Psi_{m}^{FB} \vert \Psi_{m}^{FB} \rangle \vert^{2}.
\label{qmbis}
\end{eqnarray}
This expression gives a natural geometrical interpretation of the FB conductivity. Note that one can derive another useful expression of the FB conductivity. We start with the alternative definition of the current operator, $\widehat{v}_{x}=-\frac{i}{\hbar}\left[ \widehat{x} ,\widehat{H} \right]$. This leads to the useful relation,
\begin{eqnarray}
\langle \Psi_{j}^{FB} \vert  \widehat{v}_{x} \vert \Phi_l \rangle = -i \langle \Psi_{j}^{FB} \vert  \widehat{x} \vert \Phi_l \rangle \bar{E_l}/\hbar.
\end{eqnarray}
Then, we insert this matrix element in the Kubo-Greenwood formula as it is given in the main text and we obtain,
\begin{eqnarray}
\sigma_{fb}= \frac{2\sigma_{0} }{N_c} \sum_{j} \left[ \langle \Psi_{j}^{FB} \vert \widehat{x}^{2} \vert \Psi_{j}^{FB} \rangle - \sum_{k}\vert \langle  \Psi_{j}^{FB} \vert \widehat{x} \vert \Psi_{k}^{FB} \rangle \vert^{2} \right]. \nonumber \\
\end{eqnarray}
In the second part of the sum on the right side, we keep only $\langle \Psi_{j}^{FB} \vert \widehat{x} \vert \Psi_{j}^{FB} \rangle$, the other non vanishing overlaps ($k=j\pm 1$) can be neglected in the dilute limit as discussed above. Thus, we get,
\begin{eqnarray}
\sigma_{fb}= 2 y \, \langle L^{2}_{m} \rangle \,\sigma_0,
\end{eqnarray}
$L_{m}= \left[ \langle \Psi_{m}^{FB} \vert (\hat{x}^{2} - \langle \hat{x} \rangle^{2}) \vert \Psi_{m}^{FB} \rangle \right]^{1/2}$ is a measure of the mean spread of the FB eigenstate $\vert \Psi_{m}^{FB} \rangle$.
The relation between the quantum metric and the spread of the eigenstates has been introduced in the context of maximally localized Wannier \cite{marzari1,marzari2}. 

We proceed further and calculate the value of $g^{m}_{xx}$ associated to the flat band state $\vert \psi_{0m}\rangle $. From the expression of the FB eigenstate as it is given in Eq.(\ref{eqfb1}), one finds,
\begin{eqnarray}
\langle \partial_{\varphi} \Psi_{m}^{FB} \vert \partial_{\varphi} \Psi_{m}^{FB} \rangle &=& \frac{1}{2m+1}\left( m(m-2) + 
2\sum_{1}^{m-1} k^{2} \right) \nonumber \\ 
& \approx & \frac{1}{3}m^{2},
\end{eqnarray}
and the second term is,
\begin{eqnarray}
\langle \partial_{\varphi} \Psi_{m}^{FB} \vert \Psi_{m}^{FB} \rangle &=& \frac{i}{2m+1} \left( m-2+2\sum_{1}^{m-1} k \right) \nonumber \\ 
 &\approx& i\frac{m}{2}.
\end{eqnarray}
Then, the quantum metric associated to the disordered FB eigenstate $\vert \Psi_{m}^{FB} \rangle$ has the simple form,
\begin{eqnarray}
g^{m}_{\varphi \varphi}=\frac{m^{2}}{12} + o(1).
\label{gm}
\end{eqnarray}
For a given configuration of disorder (random position of the $y \cdot N_c$ atoms of B type), the probability that the distance (in $a$) between two successive B atoms is $l$ can be approximated in the dilute limit by a Poisson distribution, $P_y(l)=ye^{-yl}$. Using the expression of the quantum metric as it is given in Eq.(\ref{gm}), we immediately find,
\begin{eqnarray}
\langle g^{m}_{\varphi \varphi}\rangle = \frac{1}{6y^{2}}.
\end{eqnarray}
Thus, in the disordered sawtooth chain the conductivity at the FB energy is,
\begin{eqnarray}
\sigma_{fb}=\frac{\sigma_{0}}{3y}. 
\label{sigdd}
\end{eqnarray}
In order to evaluate the impact of the disorder, one can straightforwardly calculate the conductivity in the case where the B atoms are now organized on a superlattice. The distance between B atoms is constant, $l=\bar{l}=1/y$. In this case the probability distribution reduces to $P_y(l)=\delta(l-\bar{l})$. This immediately leads to $\langle g^{m}_{\varphi \varphi}\rangle = \frac{1}{12y^{2}}$. Hence, in the ordered case the average quantum metric is half of that of the disordered system. Thus the conductivity is twice as large in the former case than in the later one. The disorder enhances the FB conductivity.

It is also interesting to calculate the quantum metric in the clean sawtooth chain. In this case, the exact FB eigenstates are,
$\vert \Psi_{k}^{FB} \rangle=\frac{1}{\sqrt{2+\cos(ka)}} \left[\vert A,k\rangle -\sqrt{2} \cos(ka/2)e^{-ika/2} \vert B,k \rangle \right] $ where $k$ is the momentum and $\vert X,k\rangle =\frac{1}{\sqrt{N_c}} \sum_{i} e^{ikR_i}\vert X,i \rangle$ ($X=A,B$)
From Eq.(\ref{qmbis}) where $\partial_{\varphi}=\partial_{ka}$ one finds after some steps, the quantum metric,
\begin{eqnarray}
g^{k}_{\varphi \varphi}=\frac{1}{2(2+\cos(ka))^{2}}. 
\label{gmb}
\end{eqnarray}
Then, we find $\langle g^{k}_{\varphi \varphi}\rangle=\frac{1}{2\pi} \int_{0} ^{\pi}\frac{1}{(2+\cos(ka))^{2}} dk= \frac{1}{3\sqrt{3}}$ where the integral can be calculated exactly using the standard residue theorem. From Eq.(\ref{eqsigc}), we finally find the conductivity in the clean sawtooth chain,
$\sigma_{fb}= \frac{2}{3\sqrt{3}} \sigma_{0}  \approx 0.385\, \sigma_{0}$.

It is interesting to compare this value, with the analytical expression given in Eq.(\ref{sigdd}). They differ by 13\% only. This is surprising since Eq.(\ref{sigdd}) is valid only for $y \ll 1$.

\subsection*{Appendix C : FB conductivity and quantum metric in the stub lattice}

\renewcommand{\theequation}{C.\arabic{equation}}

\setcounter{equation}{0}

In this section we calculate the quantum metric and the FB conductivity in the disordered stub lattice.
We follow the procedure used in the case of the disodered sawtooth chain (previous sections) to construct the basis of CLS FB-states in the disordered stub lattice. 
We consider a pair of B atoms located in the first and $(m+1)th$ unit cells, with no B atoms in between. The corresponding normalized FB eigenstate reads,
\begin{eqnarray}
\vert \psi_{1,m+1}\rangle &=& \frac{e^{-i \varphi /2}}{\sqrt{m\alpha^{2}+2}} (\vert B_1  \rangle -(-1)^{m}e^{im\varphi} \vert B_{m+1} \rangle) \nonumber \\
&+& \frac{\alpha}{\sqrt{m\alpha^{2}+2}} \sum_{k=1}^{m} (-1)^{k} e^{i(k-1)\varphi} \vert C_{k} \rangle. 
\label{eqfb3b}
\end{eqnarray}
The first term of quantum metric associated to this FB eigenstate as it is defined in Eq.(\ref{qmbis}) is given by,
\begin{eqnarray}
\langle \partial_{\varphi} \Psi_{m}^{FB} \vert \partial_{\varphi} \Psi_{m}^{FB} \rangle=
\frac{1}{m\alpha^{2}+2} \sum^{3}_{i=0} f_i m^{i},
\end{eqnarray}
with $f_3= \frac{\alpha^{2}}{3}$, $f_2=(1-\frac{\alpha^{2}}{2})$, $f_1=(1+\frac{\alpha^{2}}{6})$ and $f_0=\frac{1}{2}$.
The second term in Eq.(\ref{qmbis}) is,
\begin{eqnarray}
\vert \langle \partial_{\varphi} \Psi_{m}^{FB} \vert \Psi_{m}^{FB} \rangle \vert^{2})=\frac{m^{2}}{4} \left( 1-\frac{\alpha^{2}}{m\alpha^{2}+2} \right)^{2}.
\end{eqnarray}
 The calculation of the average value of the quantum metric for a given pair $(\alpha,y)$ requires the calculations of integrals of the form $I_{np}(\alpha,y)= \int_{0}^{\infty} P_y(x)\frac{x^{n}}{(x\alpha^{2}+2)^{p}}$ where the probability distribution $P_y(x)=ye^{-yx}$, and $n$ can be $0,1,2,3$ and $p$ is $0,1, 2$. There are two limiting cases where the quantum metric can be simplified. First, we consider the case where $\alpha^{2} \ll y$, this corresponds to $m \alpha^{2} \ll 1$. One finds,
\begin{eqnarray}
g^{m}_{\varphi \varphi} = \frac{1}{4}(m+1)^{2}.
\end{eqnarray}
Thus, from eq.(\ref{eqsigc}), we obtain,
\begin{eqnarray}
\sigma_{fb}= \frac{1}{y}(1+y+ o(y^{2}))\sigma_{0}.
\end{eqnarray}
The other regime corresponds to $\alpha^{2} \gg y$ (or $m \alpha^{2} \gg 1$) for which the quantum metric reduces to $g^{m}_{\varphi \varphi} = \frac{1}{12}m^{2}$. This leads to the FB conductivity,
\begin{eqnarray}
\sigma_{fb}=\frac{1}{3y} (1+ o(y/\alpha^{2}))\sigma_{0}.
\end{eqnarray}
It is important to stress that to be valid our analytical approach requires a small overlap between the constructed FB states, which corresponds to the later regime. 

In order to estimate the impact of disorder, it is interesting to consider the case where the B atoms are organized in a superlattice, the nearest neighbour distance between B atoms is $\bar{l}=1/y$. For $\alpha^{2} \ll y$, we find $\sigma_{fb}=\frac{1}{6y} \sigma_{0}$ which is half of that of the disordered system as it has been found for the sawtooth chain.

Finally, if we consider the case of the clean stub lattice ($y=1$) we can get an exact analytical expression.
The FB eigenstates are $\vert \Psi_{k}^{FB} \rangle=\frac{1}{\sqrt{D}} \left[ \vert C,k\rangle -\frac{2\cos(ka/2)}{\alpha}\vert B,k \rangle \right] $ 
where $k$ is the momentum and $D=1+\frac{4}{\alpha^{2}}\cos^{2}(ka/2)$.
This leads to $g^{k}_{\varphi \varphi}=\frac{\sin^{2}(ka/2)}{\alpha^{2}D}$. The average of the quantum metric is,
\begin{eqnarray}
\langle g^{k}_{\varphi \varphi}\rangle=\frac{\alpha^{2}}{8\pi} \int_{0}^{\pi} \dfrac{1-\cos(k)}{(\beta+\cos(k))^{2}}dk,
\end{eqnarray}
where $\beta=1+\alpha^{2}/2$. The integral on the right hand side can be calculated exactly leading to,
\begin{eqnarray}
\langle g^{k}_{\varphi \varphi}\rangle=\dfrac{1}{2\vert \alpha \vert (4+\alpha^{2})^{1/2}}.
\end{eqnarray}

Thus, in the limit of small values of $\alpha$ the FB conductivity in the clean stub lattice is,
\begin{eqnarray}
\sigma_{fb}=\frac{1}{2\vert \alpha \vert} \sigma_{0}.
\end{eqnarray}

\end{document}